\documentclass[prb,aps,twocolumn,showpacs]{revtex4}
\usepackage{graphicx}
\usepackage[activeacute,english]{babel}                 
\usepackage{amsmath}\usepackage{amssymb}
\usepackage{amsfonts}

\begin{document}
\renewcommand{\figurename}{Fig.}
   
\title{Can one identify non-equilibrium in a three-state system by analyzing two-state trajectories?}
  
\author{Christian P. Amann, Tim Schmiedl, and Udo Seifert}

\affiliation{\mbox{II}. Institut f\"{u}r Theoretische Physik, Universit\"{a}t Stuttgart, 70550 Stuttgart, Germany} 


\begin{abstract}
For a three-state Markov system in a stationary state, we discuss whether, on the basis of data obtained from effective two-state (or \textit{on}-\textit{off}) trajectories, it is possible to discriminate between an equilibrium state and a non-equilibrium steady state. By calculating the full phase diagram we identify a large region where such data will be consistent only with non-equilibrium conditions. This regime is considerably larger than the region with oscillatory relaxation, which has previously been identified as a sufficient criterion for non-equilibrium.
\end{abstract}


\maketitle

 For the emerging field of thermodynamics of small systems, or stochastic thermodynamics \citep{bust05, seif07, rito08}, enzymes and proteins like molecular motors constitute paradigmatic systems since their conformational changes can be observed  on a single molecule level using a variety of experimental techniques \citep{selv07}. The analysis of trajectories recording the transition between microstates of the enzyme can reveal insights into both the underlying kinetics and the thermodynamic implications of such changes \citep{lu98, fish01, qian02a, seif05, min05, schm06a, gasp07, liep09}. If the surrounding conditions like temperature or concentrations of other reactants are constant, the enzyme reaches a stationary state for which two classes must be distinguished. First, in a genuine equilibrium state, no net flux between any two microstates of the enzyme occurs. Second, in a non-equilibrium steady state (NESS), the probability to find the enzyme in any microstate is still time-independent. However, non-zero net fluxes between microstates indicate that the system is externally driven with concomitant entropy production. If all microstates are experimentally accessible, in principle, distinguishing between these two alternatives is trivial, given long enough trajectories. It suffices to infer both the stationary probabilities $p^s_i$ for the system to be in state $i$ and the rates $k_{ij}$ for a transition between $i$ and $j$, and then to check for genuine equilibrium in which the detailed balance condition $p^s_i k_{ij}=p^s_j k_{ji}$ for any two states must hold.

 In practice, rarely all relevant microstates are experimentally accessible or distinguishable.\ The resulting only coarse-grained information affects and aggravates a thermodynamic consistent analysis of individual trajectories \citep{raha07,li08}. In standard applications, often only a two-state trajectory is obtained in which each observable state may in fact contain several microstates. In an ambitious program, Flomenbom and co-workers \citep{flom05, flom05a, flom06, flom08, flom08a} have investigated which information about the underlying topology of the state space, or `network', can be inferred from the observation of such two-state trajectories. But even if the underlying topology is known, two-state trajectories cannot reveal the full information like the value of all rates between microstates.

 For time-independent external conditions, arguably, the most relevant question is, whether or not such an enzyme acts in an equilibrium state or in a NESS. The simplest system for which this problem can be posed is a three-state system like the stochastic Michaelis-Menten scheme for an enzyme turning substrate into a product \citep{qian02a,rao03,kou05}. Suppose that two of the three states are experimentally not distinguishable, can we decide in which of the two classes of stationary states the enzyme is in, if we are given a sufficiently long two-state trajectory? It is well-known that oscillations in the correlation function of such a two-state trajectory imply that the underlying three-state system is in a NESS rather than in equilibrium \citep{qian00}.\ Here, we show that the range of ascertained non-equilibrium can be extended much beyond this oscillatory regime, i.e., even for decaying correlations it can be possible to infer an underlying NESS state. However, there will remain parameters for which the question posed in the title cannot be answered affirmatively. Only if all three states are visible, like in dual-color flourescence correlation spectroscopy, one can discriminate between equilibrium and non-equilibrium steady states \citep{schw97,qian04}.

 We consider a three-state Markov system with states $i=1,2,3$ and transition rates $k_{ij}$. The dynamics of the probability $p_i(t)$ to be in state $i$ at time $t$ then is governed by the master equation
 \begin{equation}
  \partial_t p_i = \sum^3_{j=1} K_{ij} p_j \label{eq:master}
 \end{equation}
 with rate matrix
 \begin{equation}
  K = \left (  \begin{array}{ccc} -k_{12}-k_{13}&k_{21}&k_{31} \\ k_{12}&-k_{21}-k_{23}&k_{32} \\ k_{13}&k_{23}&-k_{31}-k_{32} \end{array} \right). \label{eq:K}
 \end{equation}

\begin{figure}[ht]
  \includegraphics[height=0.4 \linewidth]{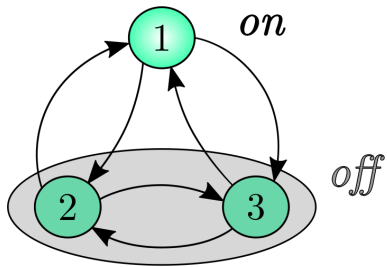}
  \caption{Three-state Markov model with states $2$ and $3$ lumped into a compound state.
  \label{fig:model}}
\end{figure}

 We assume that state $1$ (the \textit{on}-state) is fluorescently labelled and can be observed, but that it cannot be determined experimentally, whether the system is in state $2$ or state $3$ (together called the \textit{off}-state), as sketched in fig.~\ref{fig:model}. The experiment then yields an effective two-state trajectory which no longer follows a Markovian dynamics.

 Four quantities can be extracted from such two-state trajectories and identified with their theoretical correspondences as follows:

 (i) The waiting time distribution in state $1$ is exponential with a decay rate 
 \begin{equation}
  L \equiv (k_{12} + k_{13}) \label{eq:defL}.
 \end{equation}

 (ii) The waiting time distribution $w_{\rm off}(t)$ in the \textit{off}-state can be calculated from a modified Markov model where state $1$ is treated as an absorbing state which amounts to setting the rate constants $k_{12}$ and $k_{13}$ to zero. This yields the effective rate matrix
 \begin{equation}
  \widetilde{K} = \left (  \begin{array}{cc} -k_{21}-k_{23}&k_{32} \\ k_{23}&-k_{31}-k_{32} \end{array} \right),
 \end{equation}
 which governs the time evolution of the probabilities $\widetilde p_2(t)$ and $\widetilde p_3(t)$ to be in state $2$ and $3$, respectively, if we stop the dynamics when the system reaches state $1$. The waiting time distribution then is the negative time derivative of the compound probability to be either in state $2$ or state $3$,
 \begin{equation}
  w_{\rm off}(t) = - \frac {d} {dt} \left(\widetilde p_2(t) + \widetilde p_3(t) \right). \label{eq:w_off}
 \end{equation}
 This waiting time distribution is a sum of exponentials with exponents given by the eigenvalues $\tilde \lambda_i$ of the modified rate matrix $\widetilde{K}$ which can be calculated as
 \begin{equation}
  \tilde \lambda_i = \frac 1 2 \left ( -S \pm \sqrt{S^2-T} \right) \label{eq:til_la_i}
 \end{equation}
 with
 \begin{eqnarray}
  S &\equiv& k_{21} + k_{23} + k_{31} + k_{32} \quad \text{and} \label{eq:defS}\\
  T &\equiv& 4(k_{21}k_{31}+k_{21}k_{32}+k_{23}k_{31}). \label{eq:defT}
 \end{eqnarray}
 If both decay times can be determined from experimental time traces, both quantities $S$ and $T$ can be inferred.

 (iii) Last, one can measure the time-dependent probability $p_1(t)$ to be in state $1$ with initial condition $p_1(0)=1$. This quantity can also be calculated from the master equation \eqref{eq:master}, yielding a sum of exponentials
  \begin{equation}
   p_1(t) = p_1^s + c_1 e^{\lambda_1 t} + c_2 e^{\lambda_2 t}. \label{eq:p1}
  \end{equation}
 The exponents $\lambda_i$ are eigenvalues of the rate matrix $K$ \eqref{eq:K} and can be written as
  \begin{equation}
   \lambda_i = \frac 1 2 \left ( -L -S \pm \sqrt{(S+L)^2-T-M} \right) \label{eq:la_i}
  \end{equation}
 with
  \begin{equation}\begin{aligned}
   M \equiv\ &4 k_{12} (k_{23} + k_{31} + k_{32}) \\
       + &4 k_{13} (k_{21} + k_{23} + k_{32}). \label{eq:defM}
  \end{aligned}\end{equation}
 The prefactors $c_{1,2}$ can be obtained from the linear combination of eigenvectors of the matrix $K$ which yield the initial state $p_1(0)=1$. 

 In summary, the four quantities $L,\ S,\ T$ and $M$ can be determined from experimental time traces, provided the statistics is good enough to allow fits parametrizing the given exponential behavior. In principle, one could derive some other set of four independent quantities rather than $L,\ S,\ T$ and $M$. Our choice is convenient due to two reasons. First, all four quantities are non-negative (even positive if no rate constant $k_{ij}$ vanishes). Second, for the distinction between equilibrium and a NESS the absolute time-scale is irrelevant. This fact can be used to scale all times by $1/L$ leading to the three dimensionless quantities
  \begin{equation}
   \widehat S \equiv S/L, \quad \widehat T \equiv T/L^2 \quad \text{and} \quad \widehat M \equiv M/L^2.
  \end{equation}
 A rescaling of time in the original model would lead to five independent rate constants $\widehat k_{ij} \equiv k_{ij}/L$ (with $\widehat k_{12} + \widehat k_{13} = 1$). Hence given measured reduced rates $\widehat S,\ \widehat T$ and $\widehat M$ there still remains a two-parameter manifold in the three-state system dynamics that is consistent with the observed two-state trajectory.

\begin{figure*}[H,floatfix]
  \centering\includegraphics{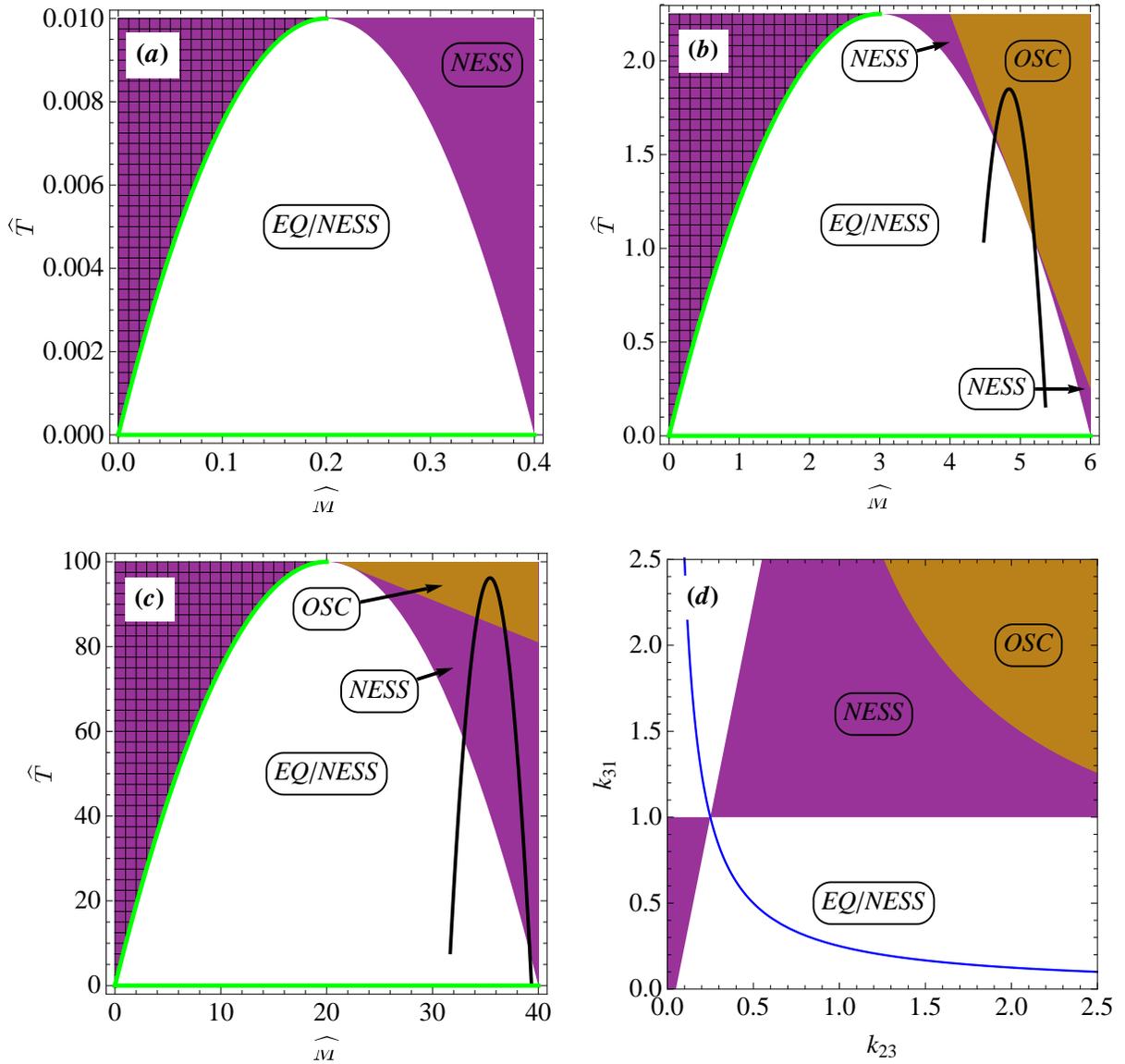}
  \caption{(color online) (a)-(c): Phase diagram in the $(\widehat M,\,\widehat T)$ plane for different values of $\widehat S$ with (a) $\widehat S=0.1$, (b) $\widehat S=1.5$ and (c) $\widehat S=10$. The white area \textit{EQ/NESS} allows both equilibrium states and NESSs, the purple area \textit{NESS} can contain only NESSs and the brown area \textit{OSC} corresponds to NESSs with oscillatory correlation function. The \textit{OSC} regime touches the \textit{EQ/NESS} regime at points $\left(2 + 2 \widehat S,-1 + \widehat S^2\right)$. The checkered area is excluded since no choice of non-negative rate constants will lead to such an $(\widehat M,\,\widehat T)$ value. The green line represents equilibrium states with some of the rate constants equal to zero. The black lines are the projection of the subdomain of the phase diagram shown in (d) that corresponds to the given $\widehat S$ value.\newline (d): Phase diagram in the $(k_{23},\,k_{31})$ plane for fixed rate constants $k_{13} = 1$, $k_{21} = 1$, $k_{32} = 1$ and $k_{12} = 4$. For each point in the diagram, the parameters $L,\ S,\ T$ and $M$ are calculated and the three regimes accordingly identified. The blue hyperbola comprises the genuine equilibrium state according to eq.~\eqref{eq:eqcon}. \label{fig:phases_in_STM_space}}
 \end{figure*}

 The three-state system is in equilibrium if and only if the original rates obey the condition
 \begin{equation}
  k_{12} k_{23} k_{31} = k_{32} k_{21} k_{13}. \label{eq:eqcon}
 \end{equation}
 An analysis of the functional dependence of the rates $k_{ij}$ on $\widehat S,\ \widehat T\ ,\widehat M$ and the equilibrium condition \eqref{eq:eqcon} assisted by Wolfram's algebraic software package M{\footnotesize ATHEMATICA}~7 leads to the identification of three regimes in the phase diagram shown in fig.~\ref{fig:phases_in_STM_space}. It is checked whether or not for given reduced rates $\widehat S$ and $\widehat M$ positive rate constants $k_{ij}$ compatible with \eqref{eq:eqcon} can be found. This is the case only if 
 \begin{equation}
  0 < \widehat T \leqslant \widehat T_1 \equiv \widehat M \widehat S - \frac{\widehat M^2}{4}, \label{eq:phases}
 \end{equation}
 which defines the regime denoted \textit{EQ/NESS} in fig.~\ref{fig:phases_in_STM_space}. In this regime, the measured data is principally insufficient for discriminating a NESS from an equilibrium state.

 Hence for any measured $\widehat T$ with $\widehat T > \widehat T_1$ the system is definitely in a NESS. This criterion constitutes our main result. We now show that this parameter space is much larger than the set of parameters for which oscillations in the correlation function occur. The latter property has previously been identified as a sufficient criterion for a NESS. Oscillations in the autocorrelation function $p_1(t)$ \eqref{eq:p1} occur if the decay rates $\lambda_i$ in eq.~\eqref{eq:la_i} acquire an imaginary part, i.e., if
 \begin{equation}
  \widehat T > \widehat T_2 \equiv (\widehat S + 1)^2 - \widehat M. \label{eq:osziphases}
 \end{equation}
 This regime is denoted by \textit{OSC} in fig.~\ref{fig:phases_in_STM_space}. It is easy to check that $\widehat T_2(\widehat M) \geqslant \widehat T_1(\widehat M)$ (with equality only for $\widehat M=2(\widehat S + 1)$). For all measured $\widehat T$ with  $\widehat T_2 \geqslant \widehat T > \widehat T_1$, denoted by \textit{NESS} in fig.~\ref{fig:phases_in_STM_space}, our new criterion establishes a NESS on the basis of two-state trajectories which do not exhibit oscillatory behaviour. While for $\widehat S \simeq 1$ the newly established region of NESSs beyond the oscillatory regime is rather small, see fig.~\ref{fig:phases_in_STM_space}(b), both for $\widehat S \ll 1$ and for $\widehat S \gg 1$ the present analysis shows a large regime of newly ascertained NESSs. In the latter two cases, the oscillatory regime is, for $\widehat S \leqslant 0.5$, totally absent or rather small, see figs.~\ref{fig:phases_in_STM_space}(a) and (c), respectively.

 For fixed $\widehat S$ the phase diagrams of physically admissible $(\widehat M,\,\widehat T)$ values are restricted to $\widehat M \leqslant 4\widehat S$ and $\widehat T \leqslant \widehat S^2$, since values outside this range would require negative $k_{ij}$ values. Moreover, for $\widehat M \leqslant 2\widehat S$, only $T \leqslant \widehat T_1$ can be realized with non-negative $k_{ij}$. The region $\widehat S^2 \geqslant T > \widehat T_1$, which is excluded by the latter condition, is shown checkered in figs.~\ref{fig:phases_in_STM_space}(a)-(c). The equalities in these constraints of physically admissible values hold only if some of the rates are zero.

 Finally, an analysis of the boundary $\widehat T_1(\widehat M)$, which requires some rate constants $k_{ij}$ to be zero, shows that for $0 \leqslant \widehat M \leqslant 2 \widehat S$ the system is definitely in equilibrium on this boundary. For $\widehat T = 0$ the system is also definitely in equilibrium.

 For representative examples of these results in the original space of the six rates $k_{ij}$, consider fig.~\ref{fig:phases_in_STM_space}(d). There we show the three regimes \textit{EQ/NESS}, \textit{NESS} and \textit{OSC} in the ($k_{23}$, $k_{31}$) plane holding the other four rate constants fixed. The line of genuine equilibrium \eqref{eq:eqcon} is embedded into region \textit{EQ/NESS} where information from the state trajectory is not sufficient to uniquely identify equilibrium. However for ($k_{23}$, $k_{31}$) parameter values in the region \textit{NESS} our analysis predicts correctly a NESS beyond the previously established \textit{OSC} region.

 Experimental measurements of the waiting time distribution $w_{\rm off}(t)$~\eqref{eq:w_off} and autocorrelation function $p_1(t)$~\eqref{eq:p1} will cause some uncertainty due to finite data. For a rough estimate, we assume a relative error of $\pm\varepsilon$ in all eigenvalues $\tilde \lambda_i$~\eqref{eq:til_la_i} and $\lambda_i$~\eqref{eq:la_i}, while it should be possible to determine the decay time $L$~\eqref{eq:defL} with better precision. A linear error propagation then yields a constant relative error for $\widehat S$ and $\widehat T$ of $\pm\varepsilon$ and $\pm2\varepsilon$ respectively. The effect on $\widehat M$ is a relative error of $\pm(2\varepsilon+4\varepsilon\,\widehat T/\widehat M)$. Thus error bars of experimentally determined pairs of $(\widehat M,\widehat T)$ values would increase linearly on straight lines through the origin in fig.~\ref{fig:phases_in_STM_space}(a)-(c). Since the maximum value of $\widehat T$ increases quadratically with the maximum value of $\widehat M$, the uncertainty on the phase boundary between \textit{NESS} and \textit{EQ/NESS} in fig.~\ref{fig:phases_in_STM_space}(a)-(c) increases with increasing $\widehat S$. Hence, a clear discrimination between the two regions would be affected by errors less in the small $\widehat S$ regime than for large $\widehat S$. For the data shown in fig.~\ref{fig:phases_in_STM_space}(a) and $\varepsilon=0.05$, the phase boundary effectively broadens to about $11\%$ on the top and $10\%$ at the bottom.\\
 \indent In summary, we have derived a criterion sufficient to identify a non-equilibrium steady state in a three-state system based on information extracted from sufficently long two-state trajectories. While the new criterion extends the region of ascertained NESSs significantly beyond the previously identified region based on oscillation in the autocorrelation function, there remain parameter values for which two-state trajectories contain principally not enough information to discriminate NESS from an equilibrium state. It will be interesting to perform a similar analysis for four-state systems or even more complex networks, where one would expect that the region of ascertained NESSs on the basis of two-state trajectory information becomes relatively smaller than in the three-state case. Generalization of our approach to cases where three-state trajectories, obtained, e.g., via dual-colour fluorescence, are available on four or more-state networks is feasible in principle, but may, in practice, become algorithmically challenging.

\bibliographystyle{apsrev}

\end{document}